\documentclass{article}

\usepackage{arxiv}

\usepackage[utf8]{inputenc} 
\usepackage[T1]{fontenc}    
\usepackage{hyperref}       
\usepackage{url}            
\usepackage{booktabs}       
\usepackage{amsfonts}       
\usepackage{nicefrac}       
\usepackage{microtype}      
\usepackage{lipsum}		
\usepackage{graphicx}
\usepackage[english]{babel}
\usepackage{doi}

\usepackage{algorithm}
\usepackage{amsmath} 
\usepackage[capitalize]{cleveref} 

\newcommand{\lword}[1]{\leavevmode\nobreak\hskip0pt plus\linewidth\penalty50\hskip0pt plus-\linewidth\nobreak#1}

\usepackage{quantikz} 
\usepackage{makecell} 
\usepackage{bm}

\title{Quantum Dynamic Time Warping for Multivariate Time Series Classification}

\author{ 
    \href{https://orcid.org/0000-0001-5790-0577}{\includegraphics[scale=0.06]{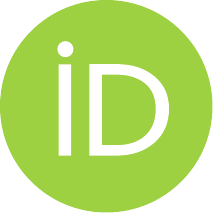}\hspace{1mm}Diego Alvarez-Estevez}\thanks{Corresponding author.} \\
	CITIC Research Center \\
	Universidade da Coruña \\
	A Coruña, 15071, Spain \\
	\texttt{diego.alvareze@udc.es} \\
	\And
    \href{https://orcid.org/0009-0007-7084-3678}{\includegraphics[scale=0.06]{orcid.pdf}\hspace{1mm}Alejandro Mayorga-Redondo} \\
	CITIC Research Center \\
	Universidade da Coruña \\
	A Coruña, 15071, Spain \\
	\texttt{alejandro.mayorga@udc.es} \\
    \And
    \href{https://orcid.org/0000-0002-4894-1067}{\includegraphics[scale=0.06]{orcid.pdf}\hspace{1mm}Eduardo Mosqueira-Rey} \\
	CITIC Research Center \\
	Universidade da Coruña \\
	A Coruña, 15071, Spain \\
	\texttt{eduardo.mosqueira@udc.es} \\
}

\date{}

\renewcommand{\headeright}{}
\renewcommand{\undertitle}{}
\renewcommand{\shorttitle}{QKT benchmarking}

\begin{document}
\maketitle

\begin{abstract}
Dynamic Time Warping (DTW) is a cornerstone for time series classification, but its reliance on Euclidean distances fails to capture latent cross-channel correlations in complex multivariate data. We propose a hybrid Quantum Dynamic Time Warping (qDTW) architecture, replacing the classical distance metric with the parameterized geometry of a quantum Hilbert space. Through structural ablation on benchmarks up to $C=8$ spatial dimensions, we establish fundamental topological rules for quantum sequence alignment.

We introduce a Unified Pre-Embedding Adjoint Ansatz that decouples trainable entanglement from classical data, eliminating the severe phase-scrambling and information bottlenecks inherent to traditional measurements. We demonstrate this decoupled architecture allows untrained quantum kernels to act as highly expressive baselines, while parameterized training effectively untangles deeply overlapping hyper-dimensional data.

Furthermore, we identify a strict spatial-temporal expressivity tradeoff: temporal depth (data re-uploading) is necessary for dimensionally restricted univariate circuits, but applying it to wide multi-qubit registers triggers chaotic frequency-spectrum explosions and representation collapse. By navigating these topological hazards, our multivariate quantum architecture outperforms classical baselines, setting a new standard for integrating parameterized quantum circuits with dynamic programming
\end{abstract}


\section{Introduction}
The classification of sequential time series is a foundational challenge in machine learning, driving applications across diverse domains from speech recognition and financial forecasting to wearable biomechanics and clinical diagnostics. Historically, approaches to this problem have diverged into several distinct methodological paradigms. Feature-based and dictionary-based algorithms extract summary statistics or discrete sub-patterns to feed into classical classifiers \cite{bagnall2017great}, while modern deep learning architectures, such as Convolutional Neural Networks (CNNs) and Long Short-Term Memory (LSTM) networks, learn hierarchical spatial-temporal representations directly from raw sequences \cite{fawaz2019deep}. However, deep learning models are notoriously data-hungry and opaque, whereas feature extraction frequently discards the holistic temporal dynamics of the signal. 

Consequently, whole-sequence distance-based classification, typically coupled with a $k$-Nearest Neighbors ($k$-NN) algorithm, remains one of the most robust, interpretable, and competitive baselines in the field. A defining characteristic of real-world temporal data is its inherent elasticity; signals frequently exhibit localized variations in speed, duration, and phase. Because rigid point-to-point distance metrics fail to capture this elasticity, Dynamic Time Warping (DTW) elegantly resolves this limitation by computing an optimal, non-linear temporal alignment \cite{sakoe1978dynamic}. Nevertheless, standard DTW algorithms predominantly utilize classical Euclidean distance to evaluate their pointwise cost matrices. While effective for simple univariate data, Euclidean space fundamentally assumes feature independence, rendering it incapable of capturing the complex, non-linear cross-channel correlations inherent to multivariate time series.

To overcome the limitations of classical Euclidean spaces, Quantum Machine Learning (QML) has emerged as a powerful alternative for geometric feature mapping. Parameterized Quantum Circuits (PQCs) leverage the exponentially large dimensionality of the quantum Hilbert space to construct highly expressive, non-linear decision boundaries via quantum embeddings \cite{schuld2019quantum, benedetti2019parameterized}. For temporal and spatial data, techniques such as data re-uploading allow continuous classical sequences to be mapped directly into the rotational phases of qubits \cite{perez2020data}, while quantum entanglement provides a native physical mechanism to bind independent data channels into a singular, correlated state. Despite these theoretical advantages, the integration of QML with dynamic programming for sequence alignment remains critically underexplored. Attempts to scale these hybrid architectures are frequently derailed by severe trainability bottlenecks. Most notably, these include the Barren Plateau problem, where gradients vanish exponentially with respect to qubit count and circuit depth \cite{mcclean2018barren}, and the related phenomenon of exponential concentration in quantum kernel methods \cite{Thanasilp_2024}.

Recent efforts to bring quantum enhancements to time-series analysis have largely pursued divergent structural strategies. For continuous forecasting, models have successfully utilized data re-uploading within Quantum Neural Networks (QNNs) \cite{schetakis2025} or employed Parameterized Quantum Circuits as fixed convolutional reservoirs \cite{strata2025}. To address temporal elasticity and sequential data, some architectures bypass explicit sequence alignment entirely, opting instead for Quantum Recurrent Neural Networks (QRNNs) to model multivariate temporal dependencies \cite{viqueira2025}, with some variants equipped with adaptive gating to achieve inherent time-warping invariance \cite{Nikoloska_2023}. When explicit Dynamic Time Warping (DTW) is addressed, the quantum literature has predominantly focused on accelerating the computational complexity of the warping path search. For instance, Feld et al. formulated the DTW alignment grid as a Quadratic Unconstrained Binary Optimization (QUBO) problem intended for Quantum Annealers \cite{feld2020}, while Zaiou (2022) explored quantum algorithmic frameworks for accelerating sequence and subsequence matching \cite{zaiou2022}. Crucially, these DTW-based approaches retain rigid classical distances (e.g., Euclidean distance) to compute the underlying cost matrix, utilizing quantum mechanics solely to optimize the dynamic programming step. 

In stark contrast, our approach addresses the representational bottleneck rather than the computational one. In this study, we introduce a Quantum DTW (qDTW) engine in which we leave the temporal warping path search to classical dynamic programming, and instead replace the naive classical Euclidean metric with a highly expressive, learnable quantum distance by embedding multivariate time steps directly into the rotational gates of a PQC. We address the optimization process through the lens of \textit{Quantum Metric Learning} \cite{lloyd2020quantum}. To ensure trainability and prevent gradient decay, the quantum embedding space is explicitly sculpted using a Triplet Margin Loss \cite{schroff2015facenet}, while temporal pathing is constrained via a Sakoe-Chiba band to accelerate computation and prevent pathological alignments \cite{sakoe1978dynamic}.

Through a rigorous evaluation of this hybrid architecture across varying spatial dimensionalities (scaling up to $C=8$) and measurement topologies, this study establishes the strict architectural requirements needed for quantum dimensionality to yield a definitive empirical advantage over classical baselines.

Our main contributions are threefold:
\begin{enumerate}
    
    \item \textbf{The Unified Pre-Embedding Ansatz and the Untrained Baseline:} We introduce a globally measured, Adjoint-based quantum metric that physically decouples trainable parameters from classical data loading. We demonstrate that this architecture successfully untangles hidden multivariate correlations, eliminating the severe phase-scrambling and information bottlenecks caused by traditional local observables ($\langle Z_0 \rangle$). Furthermore, we establish that this topological decoupling allows the \textit{untrained} quantum kernel to function as a highly expressive, deployment-ready "lazy learning" baseline that natively rivals or outperforms classical dynamic programming prior to any gradient optimization.
    
    \item \textbf{The Spatial-Temporal Expressivity Tradeoff:} We provide empirical evidence that spatial width (qubit count) and temporal depth (data re-uploading) act as mutually exclusive expressivity mechanisms in QML. While single-qubit re-uploading is strictly necessary to overcome spatial deficits in univariate data, applying deep temporal interleaving to high-dimensional multivariate registers (up to $C=8$) triggers a frequency-spectrum explosion, resulting in catastrophic representation collapse and optimization paralysis.
    
    \item \textbf{Triplet-Optimized Alignment and Architectural Limits:} We formalize a novel methodology for training quantum dynamic programming matrices using Triplet Margin Loss, proving that parameterized quantum states require strictly relative boundaries to prevent phase-wrapping collisions. Concurrently, we identify a strict architectural threshold at a circuit depth of $L=3$ in our tested data, achieving the precise spatial expressivity required for complex multi-channel data while actively preventing the onset of barren plateaus in deeper architectures.
\end{enumerate}

By exposing both the distinct mathematical advantages of the global quantum metric and the strict topological limits of parameterized entanglement, this paper provides a robust, empirically grounded framework for the future application of Quantum Dynamic Time Warping.

\section{Methods}
\label{sec:methods}

To evaluate the efficacy of quantum-enhanced distance metrics, we developed a hybrid quantum-classical architecture integrating Parameterized Quantum Circuits (PQCs) with dynamic programming. Framing our approach within the context of quantum metric learning \cite{lloyd2020quantum}, we replace the classical pointwise Euclidean distance with a trainable quantum distance metric based on state fidelity \cite{havlicek2019supervised, schuld2019quantum}, optimized to dynamically separate sequence classes within the quantum Hilbert space.

\subsection{Quantum Dynamic Time Warping (qDTW)}

Let a time series sequence be defined as $X = \{x_1, x_2, \dots, x_T\}$, where $T$ is the sequence length. In the univariate case, each $x_t \in \mathbb{R}$. In the multivariate case, $x_t \in \mathbb{R}^C$, where $C$ is the number of distinct channels (e.g., $X, Y, Z$ accelerometer axes). 

The standard Dynamic Time Warping (DTW) algorithm was developed to measure the similarity between two temporal sequences that may vary in speed or phase \cite{sakoe1978dynamic}. Hence, given two classical sequences, an anchor $X = \{x_1, \dots, x_{T_x}\}$ and a target $Y = \{y_1, \dots, y_{T_y}\}$, DTW computes the optimal non-linear alignment by first evaluating a localized distance between all point pairs. 

In classical DTW, the point-to-point distance between time steps $x_i$ and $y_j$ is predominantly calculated using the rigid Euclidean metric: $d_C(x_i, y_j) = ||x_i - y_j||_2$. The algorithm then recursively evaluates an accumulated cost matrix $D_C$ to find the optimal warping path that minimizes the total alignment cost:

\begin{equation}
    D_C(i, j) = d_C(x_i, y_j) + \min 
    \begin{cases} 
        D_C(i-1, j) \\ 
        D_C(i, j-1) \\ 
        D_C(i-1, j-1) 
    \end{cases}
\end{equation}

\begin{figure}[h]
\centering
\resizebox{\linewidth}{!}{
\begin{tikzpicture}[scale=0.9, thick]

    \begin{scope}[xshift=0cm]
        \node[font=\large\bfseries] at (3, 7) {Euclidean Alignment};

        \coordinate (x0) at (0, 4.5);
        \coordinate (x1) at (1, 4.5);
        \coordinate (x2) at (2, 6.0); 
        \coordinate (x3) at (3, 4.0); 
        \coordinate (x4) at (4, 6.0); 
        \coordinate (x5) at (5, 4.5);
        \coordinate (x6) at (6, 4.5);

        \coordinate (y0) at (0, 1.5);
        \coordinate (y1) at (1, 1.5);
        \coordinate (y2) at (2, 1.5);
        \coordinate (y3) at (3, 3.0); 
        \coordinate (y4) at (4, 1.0); 
        \coordinate (y5) at (5, 3.0); 
        \coordinate (y6) at (6, 1.5);

        \draw[blue, very thick] (x0) -- (x1) -- (x2) -- (x3) -- (x4) -- (x5) -- (x6);
        \draw[red, very thick] (y0) -- (y1) -- (y2) -- (y3) -- (y4) -- (y5) -- (y6);

        \node[blue, left=0.2cm] at (x0) {\textbf{Sequence X}};
        \node[red, left=0.2cm] at (y0) {\textbf{Sequence Y}};

        \foreach \i in {0,...,6} {
            \draw[dashed, gray, thick] (x\i) -- (y\i);
        }

        \foreach \i in {0,...,6} {
            \filldraw[blue] (x\i) circle (2.5pt);
            \filldraw[red] (y\i) circle (2.5pt);
        }
    \end{scope}

    \begin{scope}[xshift=9.5cm]
        \node[font=\large\bfseries] at (3, 7) {DTW Alignment};

        \coordinate (dx0) at (0, 4.5);
        \coordinate (dx1) at (1, 4.5);
        \coordinate (dx2) at (2, 6.0); 
        \coordinate (dx3) at (3, 4.0); 
        \coordinate (dx4) at (4, 6.0); 
        \coordinate (dx5) at (5, 4.5);
        \coordinate (dx6) at (6, 4.5);

        \coordinate (dy0) at (0, 1.5);
        \coordinate (dy1) at (1, 1.5);
        \coordinate (dy2) at (2, 1.5);
        \coordinate (dy3) at (3, 3.0); 
        \coordinate (dy4) at (4, 1.0); 
        \coordinate (dy5) at (5, 3.0); 
        \coordinate (dy6) at (6, 1.5);

        \draw[blue, very thick] (dx0) -- (dx1) -- (dx2) -- (dx3) -- (dx4) -- (dx5) -- (dx6);
        \draw[red, very thick] (dy0) -- (dy1) -- (dy2) -- (dy3) -- (dy4) -- (dy5) -- (dy6);

        \node[blue, left=0.2cm] at (dx0) {\textbf{Sequence X}};
        \node[red, left=0.2cm] at (dy0) {\textbf{Sequence Y}};

        \draw[dashed, gray, thick] (dx0) -- (dy0);
        \draw[dashed, gray, thick] (dx1) -- (dy1);
        \draw[dashed, gray, thick] (dx1) -- (dy2); 
        \draw[dashed, gray, thick] (dx2) -- (dy3); 
        \draw[dashed, gray, thick] (dx3) -- (dy4); 
        \draw[dashed, gray, thick] (dx4) -- (dy5); 
        \draw[dashed, gray, thick] (dx5) -- (dy6); 
        \draw[dashed, gray, thick] (dx6) -- (dy6); 

        \foreach \i in {0,...,6} {
            \filldraw[blue] (dx\i) circle (2.5pt);
            \filldraw[red] (dy\i) circle (2.5pt);
        }
    \end{scope}

\end{tikzpicture}
}
\caption{Comparison of classical distance metrics on phase-shifted sequential data. \textit{(Left)} Rigid Euclidean distance strictly evaluates identical temporal indices, completely failing to match the structural peaks of the signals. \textit{(Right)} Dynamic Time Warping (DTW) allows one-to-many index matchings, successfully aligning the corresponding structural features regardless of temporal distortion.}
\label{fig:dtw_concept}
\end{figure}
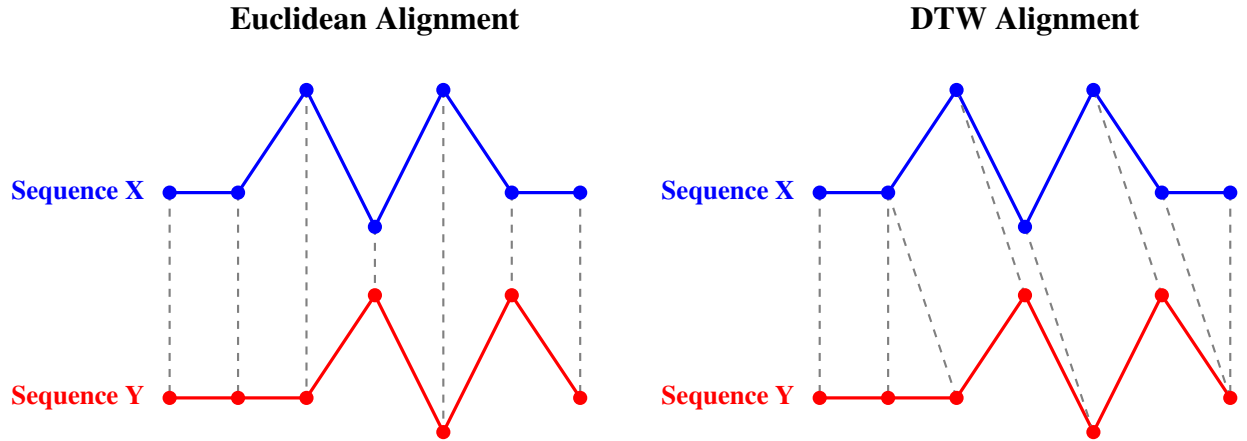

By allowing one-to-many temporal matchings, DTW successfully aligns peaks and valleys that are out of phase, as illustrated in Figure \ref{fig:dtw_concept}. However, while the dynamic programming aspect of this algorithm perfectly handles temporal elasticity, the classical Euclidean metric $d_C(x_i, y_j)$ fails to capture the complex, non-linear cross-channel correlations inherent to multivariate time series.

To resolve this representational bottleneck, we propose Quantum Dynamic Time Warping (qDTW). We completely retain the classical temporal path-finding architecture, but we substitute the classical Euclidean cost $d_C(x_i, y_j)$ with a highly expressive, trainable quantum distance metric, $d_Q(x_i, y_j)$. 

This substitution fundamentally alters the geometry of the cost matrix. While Euclidean distance computes an unbounded straight line between classical variables, our quantum distance evaluates the angular separation between temporal events mapped onto the surface of a complex Hilbert space. Specifically, we achieve this by means of using a quantum circuit that computes the fidelity $F(x_i, y_j)$ (geometric overlap) between the parameterized states encoding $x_i$ and $y_j$, measured as the expectation value of the ground state projector:

\begin{equation}
F(x_i, y_j) = |\langle \psi(x_i) | \psi(y_j) \rangle|^2.
\end{equation}

Because quantum fidelity is a measure of similarity, where $1$ indicates identical states and $0$ indicates orthogonal states, we formally define the quantum distance metric as the complement of this fidelity:
\begin{equation}
d_Q(x_i, y_j) = 1 - |\langle \psi(x_i) | \psi(y_j) \rangle|^2
\end{equation}

This formulation guarantees that identical temporal features yield a $0$ matrix cost, while maximally dissimilar steps are penalized with a distance of $1$. Consequently, the qDTW distance is naturally and tightly bounded to $[0, 1]$, providing a highly stable, normalized cost mapping for the recursive alignment algorithm:

\begin{equation}
    D_Q(i, j) = d_Q(x_i, y_j) + \min 
    \begin{cases} 
        D_Q(i-1, j) \\ 
        D_Q(i, j-1) \\ 
        D_Q(i-1, j-1) 
    \end{cases}
\end{equation}

To prevent pathological temporal alignments and to accelerate matrix construction, we enforce a Sakoe-Chiba band constraint of width $W$, restricting path calculations to $|i - j| \leq W$ \cite{sakoe1978dynamic}. In our computer simulations, pointwise quantum distances are batched and executed in a single pass prior to populating the dynamic programming matrix, mitigating sequential computational overhead (see Appendix \ref{app:implementation} for details).

\subsection{A Unified Trainable Ansatz for the Quantum Metric}

Because the qDTW matrix iteratively aligns sequences by comparing individual time steps, the underlying quantum circuit must act as a highly stable distance calculator capable of scaling from univariate ($C=1$) to high-dimensional multivariate ($C>1$) data. We designate our proposed architecture as ``Unified'' because it successfully navigates this entire dimensional spectrum using a single, consistent measurement topology, eliminating the need for dataset-specific structural redesigns.

Designing a quantum circuit capable of handling this spectrum requires navigating two severe structural bottlenecks (the empirical evidence for which is detailed in our ablation studies in Section \ref{sec:results}). First, applying deep data re-uploading \cite{perez2020data} to multi-qubit registers triggers a chaotic frequency-spectrum explosion \cite{schuld2021effect}, flattening gradients and rendering the DTW matrix untrainable. Second, placing trainable entanglement \textit{after} the classical data embedding (post-embedding) acts as a violent phase-scrambler, heavily degrading the underlying temporal geometry prior to optimization.

To bypass these limitations and compute $d_Q(x_i, y_j)$ universally, we propose a Trainable Ansatz architecture leveraging the Adjoint Fidelity measurement. To prevent unitary cancellation ($V^\dagger V = I$) and avoid barren plateaus, the trainable parameters are physically decoupled from the classical data embedding. We allocate one qubit per feature channel. The sequence overlap is evaluated via a global projector onto the ground state: $\langle 00\dots0 | V^\dagger(\theta) S^\dagger(y_j) S(x_i) V(\theta) | 00\dots0 \rangle$. 

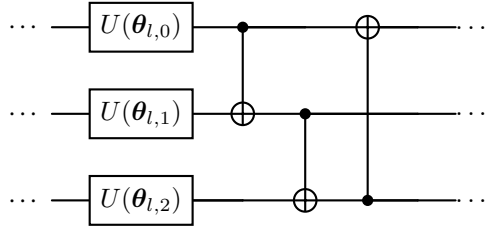
\begin{figure}[h]
\centering
\begin{quantikz}
\lstick{\dots} & \gate{U(\bm{\theta}_{l,0})} & \ctrl{1} & \qw       & \targ{}   & \qw & \dots \\
\lstick{\dots} & \gate{U(\bm{\theta}_{l,1})} & \targ{}  & \ctrl{1} & \qw       & \qw & \dots \\
\lstick{\dots} & \gate{U(\bm{\theta}_{l,2})} & \qw      & \targ{}  & \ctrl{-2} & \qw & \dots
\end{quantikz}
\caption{Internal decomposition of the $l$-th Strongly Entangling Layer of the Ansatz $V(\theta)$. The parameterized block consists of arbitrary single-qubit rotations $U(\bm{\theta}_{l,c}) = R_z(\alpha) R_y(\beta) R_z(\gamma)$ acting on each channel $c$, followed by a cascading ring of CNOT gates. This architecture generates the highly expressive spatial correlations necessary for the multivariate qDTW metric.}
\label{fig:sel_decomposition}
\end{figure}

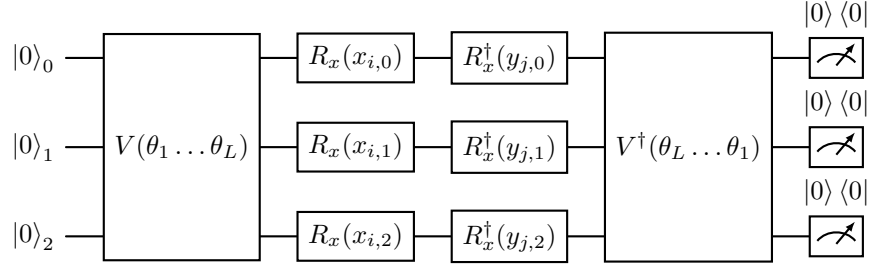
\begin{figure}[h]
\centering
\begin{quantikz}
\lstick{\ensuremath{\ket{0}_0}} & \gate[3]{V(\theta_1 \dots \theta_L)} & \gate{R_x(x_{i,0})} & \gate{R_x^\dagger(y_{j,0})} & \gate[3]{V^\dagger(\theta_L \dots \theta_1)} & \meter{\ket{0}\bra{0}} \\
\lstick{\ensuremath{\ket{0}_1}} & & \gate{R_x(x_{i,1})} & \gate{R_x^\dagger(y_{j,1})} & & \meter{\ket{0}\bra{0}} \\
\lstick{\ensuremath{\ket{0}_2}} & & \gate{R_x(x_{i,2})} & \gate{R_x^\dagger(y_{j,2})} & & \meter{\ket{0}\bra{0}}
\end{quantikz}
\caption{The Unified Pre-Embedding Adjoint Circuit computing $d_Q(x_i, y_j)$ for $C=3$. The trainable Ansatz $V(\theta)$ prepares a high-dimensional basis state prior to classical data encoding. Sequence $x_i$ is loaded forward, while sequence $y_j$ and the entanglement block are mathematically inverted. Local distance is extracted via a global geometric fidelity measurement onto the ground state $\ket{00\dots0}$.}
\label{fig:circuit_ansatz}
\end{figure}

Here, $V(\theta)$ represents a parameterized \textit{StronglyEntanglingLayer} \cite{Schuld2020_circuit_centric} of depth $L$, where the trainable parameters $\theta$ are partitioned across $L$ sequentially repeated layers of arbitrary single-qubit rotations and adjacent CNOT rings (Figure \ref{fig:sel_decomposition}). This acts as the basis state preparation prior to the classical data encoding $S(x)$, consisting of parallel $R_x$ rotations (Figure \ref{fig:circuit_ansatz}).

Because the untrained entanglement $V$ establishes a fixed high-dimensional coordinate system prior to data loading, this architecture strictly preserves relative localized geometry. Empirical results demonstrate this pre-embedding topology cures the phase-scrambling degradation seen in post-embedded architectures, allowing the untrained quantum metric to routinely match or exceed Classical DTW across all dimensionality scales, serving as a highly stable foundation for Triplet Loss optimization.

\subsection{Optimization: Triplet vs. Pairwise Margin Loss}
To optimize the quantum parameters $\theta$, we evaluated two distinct metric learning functions: Pairwise Contrastive Loss and Triplet Margin Loss \cite{schroff2015facenet}.

Pairwise Loss enforces an \textit{absolute} spatial boundary. Given a label $Y_{AB}$ (where $0$ is the same class and $1$ is a different class) and a target margin $\alpha$, it optimizes:

\begin{equation}
    \mathcal{L}_{Pair} = (1 - Y_{AB}) \cdot \text{qDTW}(A,B) + Y_{AB} \cdot \max(0, \alpha - \text{qDTW}(A,B))
\end{equation}

Conversely, Triplet Loss enforces a \textit{relative} spatial hierarchy. Given an Anchor ($A$), a Positive ($P$), and a Negative ($N$), it optimizes:

\begin{equation}
    \mathcal{L}_{Triplet} = \max(0, \text{qDTW}(A, P) - \text{qDTW}(A, N) + \alpha)
\end{equation}

While pairwise optimization is common in classical sequence alignment, enforcing absolute distances on a quantum embedding is problematic. The embedding space is defined by periodic rotational gates bounding the state vector to a closed hypersphere. Enforcing an absolute expansion ($d > \alpha$) frequently induces phase-wrapping, where pushing a negative sample "too far" causes it to wrap around the Bloch sphere, paradoxically decreasing its physical distance to the anchor. Triplet Loss mitigates this topological hazard; by enforcing a strictly relative constraint ($d(A, N) > d(A, P)$), it allows the optimizer to dynamically cluster classes anywhere on the hypersphere without violently fighting the periodic boundaries of the quantum operators.

\section{Experimental Setup}
\label{sec:setup}

\subsection{Dataset Characterization}
To evaluate the proposed qDTW architectures, we conducted experiments across four foundational time-series datasets. Three of these (BasicMotions, RacketSports, FingerMovements) are public benchmarks sourced from the UEA multivariate archive \cite{bagnall2018uea}. BasicMotions and RacketSports evaluate complex kinematic alignment, while FingerMovements is derived from magnetoencephalography (MEG) recordings and is utilized as a high-dimensional physiological stress test. To maintain computational tractability within the $O(2^C)$ simulation scaling constraints, the original 28 MEG channels of the FingerMovements dataset were deterministically truncated to the first $C=8$ spatially contiguous channels. The fourth dataset is a custom, private clinical dataset (Apnea), derived from the Sleep Heart Health Study (SHHS) \cite{quan1997sleep}, and utilized to test the architecture on real-world, unanchored physiological signals. It consists of a stratified random selection of 60-second intervals of airflow and oxygen saturation signals obtained from polysomnography (PSG) recordings, with the aim of detecting the presence or absence of sleep-disordered breathing events. To optimize the temporal length for quantum embedding, the original signals, recorded at 10 Hz and 1 Hz, respectively, were resampled to 2 Hz, resulting in a uniform sequence length of $T=120$ steps. The dataset is strictly balanced, composed of approximately 50\% apnea-hypopnea events and 50\% normal (event-free) respiratory sequences.  The characteristics of each dataset are summarized in Table \ref{tab:datasets}.

\begin{table}[h]
\centering
\caption{Characteristics of the Evaluated Datasets}
\label{tab:datasets}
\begin{tabular}{l c c c c c}
\hline
\textbf{Dataset} & \textbf{Type} & \textbf{Dimensions ($C$)} & \textbf{Length ($T$)} & \textbf{Classes} & \textbf{Total Instances} \\
\hline
BasicMotions & Kinematic (Public) & 6 & 100 & 4 & 80 \\
RacketSports & Kinematic (Public) & 6 & 30  & 4 & 303 \\
FingerMovements & Physiological/MEG (Public) & 8$^*$ & 50 & 2 & 416 \\
Apnea & Physiological (Custom) & 2 & 120 & 2 & 800 \\
\hline
\multicolumn{6}{l}{\small $^*$Deterministically truncated from 28 channels to fit an 8-qubit simulation register.}
\end{tabular}
\end{table}

\subsection{Configuration and Hyperparameters}

Prior to quantum encoding, it is a strict physical requirement to prevent global phase collisions and over-rotation within the parameterized gates. Standard Z-score normalization is dangerous in quantum circuits as unbounded outliers can cause the quantum phase to wrap chaotically. Therefore, we independently Min-Max scale every feature channel such that $x_{t, c} \in [0, \pi]$. This strict boundary mathematically anchors the classical data to the front hemisphere of the Bloch sphere, ensuring a 1-to-1 monotonic mapping between classical variables and quantum states. Following this normalization, all datasets were partitioned using a stratified split of 70\% training, 10\% validation, and 20\% testing.

All quantum circuit simulations were executed using the exact \texttt{default.qubit} state-vector simulator provided by the PennyLane quantum machine learning framework \cite{bergholm2018pennylane}. The hybrid quantum-classical gradients and loss backpropagation were managed via PyTorch \cite{paszke2019pytorch}, while matrix evaluations during the non-differentiable inference phase were accelerated using native NumPy routines \cite{harris2020array}. 

To evaluate the universal efficacy of the proposed architecture across varying dimensionalities, the Unified Pre-Embedding Adjoint Ansatz was utilized, dynamically allocating one qubit per feature channel. The basis state preparation block, $V(\theta)$, was configured with a depth of $L=3$ parameterized \textit{StronglyEntanglingLayers}. This specific depth was established as the architectural optimum based on an extensive ablation study examining $L \in \{1,3,5\}$ (detailed in Appendix A). The study confirmed that $L=3$ provides the necessary spatial expressivity to generate complex, entangled decision boundaries for multi-channel data, while remaining sufficiently shallow to prevent the catastrophic gradient attenuation (barren plateaus) and representation collapse observed at higher depths ($L=5$) \cite{mcclean2018barren}. To compute the dynamic programming matrix, a Sakoe-Chiba window size of $W=15$ was uniformly applied across all experiments to constrain the warping path and accelerate execution.

Models were trained using a batch size of 32 sequences, an initial learning rate of 0.001 with a dynamic decay scheduler, and a sampling rate of 1,000 triplets or pairs per epoch (depending on the chosen loss function). Training was capped at a maximum of 30 epochs, with an Early Stopping patience of 5 epochs to prevent over-fitting.

To optimize the quantum parameters, both Pairwise Contrastive Loss and Triplet Margin Loss were evaluated. Unlike classical Euclidean embeddings, parameterized quantum states are physically bounded by the compact geometry of the Hilbert space. Because DTW accumulates pointwise distances along a warping path, the maximum achievable distance between any two sequences is strictly constrained by the sequence length $T$. To prevent gradient saturation caused by physically unachievable margin penalties, the target margin ($\alpha$) was dynamically scaled proportional to the sequence length: $\alpha = \beta \times T$. We evaluated two base margin scalars, $\beta \in \{0.01, 0.1\}$. Consequently, shorter sequences such as \textit{RacketSports} ($T=30$) utilized a margin of $\alpha = 0.3$ at $\beta=0.01$, while extended sequences such as \textit{Apnea} ($T=120$) utilized a margin of $\alpha = 1.2$. This linear scaling ensures uniform relative clustering pressure across all varying sequence topologies.

\section{Results}
\label{sec:results}

In this section, we present the empirical evaluation of the proposed qDTW architecture. We first establish the universal performance of the Unified Pre-Embedding Ansatz, demonstrating its superiority over classical baselines across varying datasets and dimensional scaling (Section \ref{subsec:main_results}). Following this foundational evaluation, we present a series of systematic ablation studies designed to rigorously validate the topological laws governing parameterized quantum sequence alignment (Sections \ref{subsec:ablation_univariate}--\ref{subsec:ablation_measurement}).

\subsection{Universal Performance and Dimensional Scaling}
\label{subsec:main_results}

To benchmark the universal efficacy of the Unified Pre-Embedding Ansatz, performance was evaluated against a trivial Zero-R classifier (majority class prediction) to establish the mathematical floor, followed by standard Euclidean distance and classical Dynamic Time Warping (DTW). All distances were evaluated using a $k$-Nearest Neighbors ($k$-NN) classifier ($k \in \{1, 3, 5, 7, 11\}$). 

Rather than solely evaluating the datasets at their native dimensionalities, a structural scaling study was conducted across the multivariate datasets (\textit{BasicMotions}, \textit{RacketSports}, \textit{Apnea}, and \textit{FingerMovements}). The number of classical input channels was systematically increased from $C=1$ up to the native dimensionality of each dataset (reaching a maximum of $C=8$). The quantum architecture dynamically scaled the width of the qubit register to match the channel count, thereby exponentially expanding the available Hilbert space ($2^C$ dimensions). 

The results are summarized in Table \ref{tab:main_results}. Note that the reported Trained qDTW values correspond to Triplet Loss optimization using a base margin of $\beta=0.01$ and a circuit depth of $L=3$. While highly specific hyperparameter tuning can occasionally yield marginal gains on individual datasets, this configuration was selected to demonstrate the universal baseline performance of the architecture, as it was determined to be the most globally stable configuration derived from the ablation studies detailed in Appendices \ref{app:ablation_study_circuit_depth} and \ref{app:ablation_loss}.

\begin{table}[h]
\centering
\caption{Classification accuracies evaluating spatial scaling across classical channels ($C$). The Trained qDTW reflects the optimal universal configuration ($L=3$, Triplet Loss, $\beta=0.01$).}
\label{tab:main_results}
\begin{tabular}{l c c c c c c}
\hline
\textbf{Dataset} & \textbf{Channels} & \textbf{Zero-R} & \textbf{Classical Euclidean} & \textbf{Classical DTW} & \textbf{Untrained qDTW} & \textbf{Trained qDTW} \\
\hline
BasicMotions & 1 & 25.00\% & 62.50\% & 93.75\% & \textbf{100.00\%} & \textbf{100.00\%} \\
BasicMotions & 2 & 25.00\% & 68.75\% & \textbf{93.75\%} & \textbf{93.75\%} & \textbf{93.75\%} \\
BasicMotions & 3 & 25.00\% & 68.75\% & 75.00\% & 87.50\% & \textbf{100.00\%} \\
BasicMotions & 4 & 25.00\% & 62.50\% & 75.00\% & 75.00\% & \textbf{93.75\%} \\
BasicMotions & 5 & 25.00\% & 62.50\% & 75.00\% & 75.00\% & \textbf{87.50\%} \\
BasicMotions & 6 & 25.00\% & 62.50\% & 75.00\% & 75.00\% & \textbf{87.50\%} \\
\hline
RacketSports & 1 & 27.87\% & 70.49\% & \textbf{77.05\%} & 75.41\% & 75.41\% \\
RacketSports & 2 & 27.87\% & 68.85\% & \textbf{83.61\%} & 81.97\% & 80.33\% \\
RacketSports & 3 & 27.87\% & 67.21\% & \textbf{86.89\%} & 83.61\% & 83.61\% \\
RacketSports & 4 & 27.87\% & 73.77\% & 83.61\% & \textbf{85.25\%} & 81.97\% \\
RacketSports & 5 & 27.87\% & 75.41\% & \textbf{83.61\%} & \textbf{83.61\%} & \textbf{83.61\%} \\
RacketSports & 6 & 27.87\% & 75.41\% & \textbf{85.25\%} & 83.61\% & \textbf{85.25\%} \\
\hline
FingerMovements & 1 & 50.00\% & \textbf{60.71\%} & 55.95\% & \textbf{60.71\%} & \textbf{60.71\%} \\
FingerMovements & 2 & 50.00\% & 61.90\% & 61.90\% & \textbf{63.10\%} & \textbf{63.10\%} \\
FingerMovements & 3 & 50.00\% & 58.33\% & 60.71\% & 66.67\% & \textbf{67.86\%} \\
FingerMovements & 4 & 50.00\% & 55.95\% & 58.33\% & 55.95\% & \textbf{63.10\%} \\
FingerMovements & 5 & 50.00\% & 55.95\% & \textbf{64.29\%} & 60.71\% & 63.10\% \\
FingerMovements & 6 & 50.00\% & 58.33\% & 60.71\% & 60.71\% & \textbf{61.90\%} \\
FingerMovements & 7 & 50.00\% & 59.52\% & \textbf{63.10\%} & 59.52\% & 59.52\% \\
FingerMovements & 8 & 50.00\% & 54.76\% & 59.62\% & 61.90\% & \textbf{65.48\%} \\
\hline
Apnea & 1 & 50.00\% & 71.88\% & \textbf{84.38\%} & 83.75\% & 83.75\% \\
Apnea & 2 & 50.00\% & 83.75\% & 89.38\% & \textbf{92.50\%} & 90.62\% \\
\hline
\end{tabular}
\end{table}

As expected for elastic sequential data, both classical and quantum DTW versions consistently outperform both the Zero-R floor and the rigid point-to-point Classical Euclidean baseline across all datasets. This empirically confirms that non-linear temporal alignment is a fundamental requirement for these classification tasks, justifying the DTW approach. Building upon this foundation, the data demonstrates that the proposed quantum architecture routinely matches or exceeds the classical DTW baselines at full spatial integration, while exhibiting robustness when scaling to high-dimensional feature spaces.

As classical dimensionality increases, the complexity of the temporal correlations grows. For datasets such as \textit{BasicMotions}, the addition of spatial channels introduces complex kinematic overlap that degrades the Classical DTW baseline significantly (dropping from 93.75\% at $C=1$ to 75.00\% at $C=6$). While the Trained qDTW model also experiences a reduction from its perfect univariate baseline as spatial complexity increases, it successfully leverages the exponentially large Hilbert space to process these multi-channel features. By optimizing the quantum basis state preparation, the Trained qDTW maintains a highly dominant 87.50\% peak accuracy at full $C=6$ integration, sustaining a 12.5\% margin of victory over its classical counterpart.

Conversely, the \textit{RacketSports} dataset illustrates a topology where the classical DTW algorithm is already highly efficient at managing cross-channel correlations, maintaining robust accuracy as dimensionality increases. In this regime, the quantum architectures demonstrate excellent structural adaptability. At full $C=6$ integration, the Trained qDTW model successfully matches the peak Classical DTW performance of 85.25\%. This confirms that the quantum metric reliably converges to state-of-the-art performance bounds even in sequence topologies where classical dynamic programming does not catastrophically degrade.

The \textit{FingerMovements} dataset provides the most rigorous test of this scaling, expanding the evaluation up to $C=8$. This dataset exhibits high levels of cross-channel variance, evidenced by the fluctuating performance of Classical DTW. However, the optimization pipeline allows the Trained qDTW model to systematically isolate discriminative temporal patterns from these complex channel interactions, achieving the global peak accuracy for this dataset (67.86\% at $C=3$) and recovering its dominant performance margin when all 8 channels are integrated (65.48\% Trained qDTW vs. 59.62\% Classical DTW). 

Finally, the results highlight a profound structural advantage of the quantum metric: the Untrained qDTW baseline acts as a highly structural and expressive coordinate system, outright outperforming classical Euclidean metrics and rivaling or directly beating Classical DTW before a single gradient update is applied. This phenomenon is particularly evident on the \textit{Apnea} dataset (92.50\% Untrained qDTW vs. 89.38\% Classical DTW) and the highly complex 8-channel \textit{FingerMovements} dataset (61.90\% vs. 59.62\%). On \textit{RacketSports}, the Untrained qDTW baseline natively overtakes the classical model (85.25\% vs. 83.61\%) at intermediate channel capacities ($C=4$). This confirms that even parameterized random quantum circuits naturally project multivariate time series into highly separable, high-dimensional geometries.

\subsection{Ablation I: The Role of Temporal Depth in Univariate Embeddings}
\label{subsec:ablation_univariate}

While the Unified Pre-Embedding Ansatz demonstrates strong universal scaling, a single qubit ($C=1$) fundamentally lacks the spatial volume to build highly complex decision boundaries. To evaluate whether this spatial deficit can be overcome via temporal depth, we experimented with a classical data re-uploading architecture \cite{perez2020data} specifically tailored for univariate sequences. 

To calculate the distance between time steps $x_i$ and $y_j$ in this regime, we treat the classical data not as an additive angle, but as a multiplicative frequency scaling factor against trainable parameters $\theta \in [0, 1]$. The sequence $x_i$ is encoded through sequential layers of $R_y(x_i \cdot \theta_{l,1})$ and $R_z(x_i \cdot \theta_{l,2})$ gates. Bounding these parameters strictly to $[0, 1]$ prevents the multiplicatively scaled data from exceeding $\pi$, avoiding catastrophic phase wrapping. 

\begin{figure}[h]
\centering
\begin{quantikz}
\lstick{\ensuremath{\ket{0}}} & \gate{R_y(x_i \cdot \theta_1)} & \gate{R_z(x_i \cdot \theta_2)} & \dots & \gate{R_z^\dagger(y_j \cdot \theta_2)} & \gate{R_y^\dagger(y_j \cdot \theta_1)} & \meter{\ket{0}\bra{0}}
\end{quantikz}
\caption{Univariate Adjoint Re-uploading Circuit (single layer shown for brevity). The classical data acts as a phase multiplier against trainable parameters, and local distance is measured via geometric fidelity on the ground state.}
\label{fig:circuit_uni}
\end{figure}
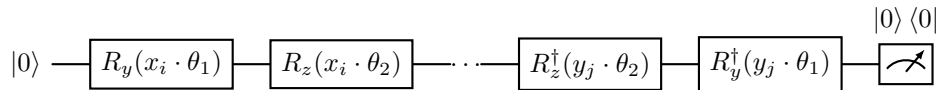

The full circuit was configured again with a depth of $L=3$ re-uploading layers, using Triplet loss and $\beta=0.01$). The distance is calculated via an adjoint overlap metric, where the inverse encoding of $y_j$ is applied to the state, terminating in a ground-state projector measurement (Figure \ref{fig:circuit_uni}). If $x_i$ and $y_j$ are identical, the forward and inverse circuits perfectly cancel, returning the qubit to $|0\rangle$ (where fidelity $\approx 1.0$).

Using this architecture, we benchmarked the univariate versions of the datasets. The results, summarized in Table \ref{tab:unidimensional_results_reup}, demonstrate the efficacy of this approach.

\begin{table}[h]
\centering
\caption{Classification accuracies using the Univariate Data Re-uploading Architecture. Results from reference Classical DTW and the proposed Pre-Embedding Ansatz are taken from Table \ref{tab:main_results}.}
\label{tab:unidimensional_results_reup}
\begin{tabular}{l c c c c c}
\hline
\textbf{Dataset} & \textbf{Channels} & \makecell{\textbf{Classical} \\ \textbf{DTW}} & \makecell{\textbf{Pre-Embedding} \\ \textbf{qDTW}} & \makecell{\textbf{Re-uploading} \\ \textbf{(Untrained)}} & \makecell{\textbf{Re-uploading} \\ \textbf{(Trained)}} \\
\hline
BasicMotions & 1 & 93.75\% & \textbf{100.00\%} & 93.75\% & \textbf{100.00\%} \\
RacketSports & 1 & \textbf{77.05\%} & 75.41\% & 75.41\% & \textbf{77.05\%} \\
FingerMovements & 1 & 55.95\% & 60.71\% & 58.33\% & \textbf{61.90\%} \\
Apnea        & 1 & \textbf{84.38\%} & 83.75\% & \textbf{84.38\%} & \textbf{84.38\%} \\
\hline
\end{tabular}
\end{table}

When compared to the baseline $C=1$ results of the single-load Pre-Embedding Ansatz (Table \ref{tab:unidimensional_results_reup}), the injection of temporal depth via data re-uploading yields distinct geometric advantages. By allowing the single qubit to iteratively twist its basis state around the classical data, the re-uploading architecture successfully elevates performance across multiple datasets. Notably, it matches the Classical DTW baselines on \textit{RacketSports} (improving from 75.41\% to 77.05\%) and \textit{Apnea} (improving from 83.75\% to 84.38\%).

More significantly, on the complex \textit{FingerMovements} dataset, the trained re-uploading architecture establishes the peak univariate accuracy of 61.90\%, outperforming both the single-load quantum baseline (60.71\%) and expanding the margin over the classical dynamic programming approach (55.95\%). This empirically confirms that when spatial width (multi-qubit entanglement) is fundamentally restricted by data dimensionality, temporal depth acts as a highly effective, complementary expressivity mechanism for quantum sequence alignment.

\subsection{Ablation II: The Detrimental Effects of Multivariate Data Re-uploading}
\label{subsec:ablation_multi_reuploading}

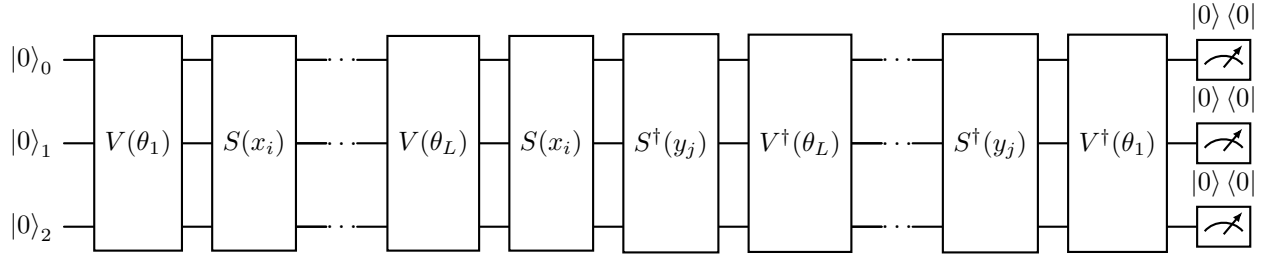
\begin{figure}[h]
\centering
\begin{quantikz}[column sep=0.4cm]
\lstick{\ensuremath{\ket{0}_0}} & \gate[3]{V(\theta_1)} & \gate[3]{S(x_i)} & \qw\dots & \gate[3]{V(\theta_L)} & \gate[3]{S(x_i)} & \gate[3]{S^\dagger(y_j)} & \gate[3]{V^\dagger(\theta_L)} & \qw\dots & \gate[3]{S^\dagger(y_j)} & \gate[3]{V^\dagger(\theta_1)} & \meter{\ket{0}\bra{0}} \\
\lstick{\ensuremath{\ket{0}_1}} & & & \qw\dots & & & & & \qw\dots & & & \meter{\ket{0}\bra{0}} \\
\lstick{\ensuremath{\ket{0}_2}} & & & \qw\dots & & & & & \qw\dots & & & \meter{\ket{0}\bra{0}}
\end{quantikz}
\caption{Multivariate Adjoint Re-uploading Circuit configured for $C=3$. The classical data encoding block $S(x)$ (comprising parallel $R_x$ rotations) and trainable entanglement block $V(\theta)$ are repeatedly interleaved $L$ times. The inverse sequence $y_j$ structurally mirrors the forward sequence $x_i$, measuring the final geometric overlap on the ground state.}
\label{fig:circuit_multi_reuploading}
\end{figure}

Having established that temporal depth is beneficial for dimensionally restricted ($C=1$) spaces, we investigated whether applying this same re-uploading architecture to multivariate sequences ($C>1$) would yield similar expressivity gains. We evaluated an $L=3$ Adjoint Re-uploading architecture (illustrated in Figure \ref{fig:circuit_multi_reuploading}) across varying spatial widths, the results of which are summarized in Table \ref{tab:results_multi_with_reuploading2}.

\begin{table}[h]
\centering
\caption{Classification accuracies using the Multivariate Adjoint Re-uploading Architecture. Results from reference Classical DTW and the proposed Pre-Embedding Ansatz are taken from Table \ref{tab:main_results}.}
\label{tab:results_multi_with_reuploading2}
\begin{tabular}{l c c c c c}
\hline
\textbf{Dataset} & \textbf{Channels} & \makecell{\textbf{Classical} \\ \textbf{DTW}} & \makecell{\textbf{Pre-Embedding} \\ \textbf{qDTW}} & \makecell{\textbf{Re-uploading} \\ \textbf{(Untrained)}} & \makecell{\textbf{Re-uploading} \\ \textbf{(Trained)}} \\
\hline
BasicMotions & 6 & 75.00\% & \textbf{87.50\%} & 75.00\% & 75.00\% \\
RacketSports & 6 & 85.25\% & 85.25\% & \textbf{88.52\%} & 86.89\% \\
FingerMovements & 8 & 59.52\% & \textbf{65.48\%} & 55.95\% & 52.38\% \\
Apnea        & 2 & 89.38\% & 90.62\% & \textbf{91.25\%} & 90.00\% \\
\hline
\end{tabular}
\end{table}

While data re-uploading is strictly necessary for univariate circuits to build sufficient spatial volume, applying temporal depth ($L>1$) to a multi-qubit register introduces severe optimization hazards. On high-dimensional datasets, the repeated interleaving of classical data and parameterized entanglement triggers a massive Fourier frequency-spectrum explosion \cite{schuld2021effect}. The resulting optimization landscape becomes highly chaotic, effectively paralyzing the optimizer.

This failure mode is observable across multiple topologies but is most catastrophic on the highly complex $C=8$ \textit{FingerMovements} dataset. While the single-load Pre-Embedding architecture successfully managed this 8-qubit space to achieve a peak accuracy of 65.48\%, transitioning to deep re-uploading dropped the untrained baseline to 55.95\%. More critically, attempting to optimize this dense frequency spectrum completely paralyzed the training pipeline, actively degrading the model's predictive power to a near-random 52.38\%.

A similar optimization stall is visible in the \textit{RacketSports} dataset ($C=6$). While the deep temporal interleaving generated a highly expressive initial geometric projection (yielding a superior untrained baseline of 88.52\%), the chaotic gradient landscape prevented the Triplet Loss from finding stable minima. Consequently, the optimizer failed to learn any meaningful geometric boundaries, triggering early stopping and degrading the trained performance down to 86.89\%. This empirically confirms that for wide multi-qubit registers, spatial width (entanglement) and temporal depth (re-uploading) act as mutually exclusive expressivity mechanisms; combining them results in fatal over-parameterization.

Further ablation on the narrow-width \textit{Apnea} dataset ($C=2$) highlights the physical threshold of this phenomenon. While reducing the spatial volume to just two qubits prevented total optimization paralysis and allowed the Triplet Loss to successfully converge, the deep $L=3$ architecture still failed to surpass its untrained baseline (dropping from 91.25\% to 90.00\%). Furthermore, across all dimensionalities, the deep adjoint architecture incurs a significant computational penalty, artificially inflating DTW matrix computation and training times compared to the single-load framework. This confirms that temporal re-uploading provides no scalable geometric or predictive advantage in multi-qubit adjoint frameworks, strictly validating the use of single-load topologies for multivariate sequence alignment.

\subsection{Ablation III: Measurement Topology and the Failure of Local Observables}
\label{subsec:ablation_measurement}

In earlier iterations of quantum machine learning, sequence alignment and classification were frequently handled via local observables (quantum pooling). To justify the use of the global Adjoint geometric overlap ($\langle \psi_A | \psi_B \rangle$) in our Unified Ansatz, we evaluated an alternative measurement topology. 

\begin{figure}[h]
\centering
\begin{quantikz}
\lstick{\ensuremath{\ket{0}_0}} & \gate{R_x(x_{i,0})} & \gate{R_y(y_{j,0})} & \gate[3]{V(\theta_1, \dots, \theta_L)} & \meter{\ensuremath{\langle Z \rangle}} \\
\lstick{\ensuremath{\ket{0}_1}} & \gate{R_x(x_{i,1})} & \gate{R_y(y_{j,1})} & & \qw \\
\lstick{\ensuremath{\ket{0}_2}} & \gate{R_x(x_{i,2})} & \gate{R_y(y_{j,2})} & & \qw
\end{quantikz}
\caption{Multivariate Local Observable Circuit for $C=3$. Both sequence steps are loaded simultaneously, passed through trainable entanglement ($V$), and measured via quantum pooling on Qubit 0.}
\label{fig:circuit_multi}
\end{figure}
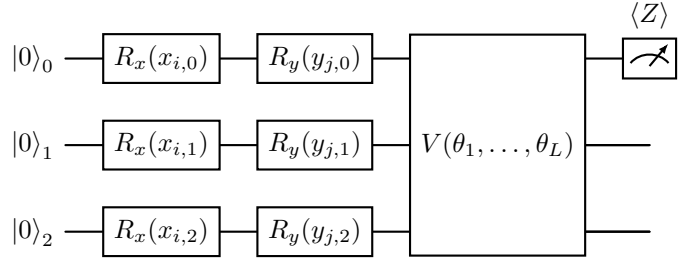

In this architecture, both the sequence point $x_i$ and the reference point $y_j$ are loaded forward onto the circuit sequentially via additive $R_x$ and $R_y$ gates. The state is then passed through a post-embedding entanglement block $V(\theta)$ (for comparison purposes, configured with $L=3$ layers). Finally, the distance is extracted by measuring the Pauli-Z expectation value of a single designated qubit, $\langle Z_0 \rangle$ (Figure \ref{fig:circuit_multi}).

The results of this ablation are summarized in Table \ref{tab:results_channels1}. To highlight the architectural tradeoff, the performance of this Local Observable circuit is directly contrasted with the peak performance of the Adjoint Pre-Embedding Ansatz (Table \ref{tab:main_results}).

\begin{table}[h]
\centering
\caption{Classification accuracies using the Local Observable ($\langle Z_0 \rangle$) architecture compared against the Adjoint Pre-Embedding Ansatz. Evaluated wit $L=3$, Triplet Loss, and $\beta=0.01$. Results from reference Classical DTW and the proposed Pre-Embedding Ansatz are taken from Table \ref{tab:main_results}.}
\label{tab:results_channels1}
\begin{tabular}{l c c c c c}
\hline
\textbf{Dataset} & \textbf{Channels} & \makecell{\textbf{Classical} \\ \textbf{DTW}} & \makecell{\textbf{Adjoint} \\ \textbf{Pre-Embedding}} & \makecell{\textbf{Local Obs.} \\ \textbf{(Untrained)}} & \makecell{\textbf{Local Obs.} \\ \textbf{(Trained)}} \\
\hline
BasicMotions & 1 & 93.75\% & \textbf{100.00\%} & 25.00\% & 75.00\% \\
BasicMotions & 6 & 75.00\% & 87.50\% & 37.50\% & \textbf{100.00\%} \\
\hline
RacketSports & 1 & \textbf{77.05\%} & 75.41\% & 24.59\% & 40.98\% \\
RacketSports & 6 & \textbf{85.25\%} & \textbf{85.25\%} & 24.59\% & 73.77\% \\
\hline
FingerMovements & 1 & 55.95\% & \textbf{60.71\%} & 48.81\% & 55.95\% \\
FingerMovements & 8 & 59.52\% & \textbf{65.48\%} & 58.33\% & 54.76\% \\
\hline
Apnea        & 1 & \textbf{84.38\%} & 83.75\% & 59.38\% & 71.25\% \\
Apnea        & 2 & 89.38\% & \textbf{90.62\%} & 50.00\% & 65.00\% \\
\hline
\end{tabular}
\end{table}

This ablation reveals two fatal flaws inherent to post-embedding local measurement topologies. First, applying parameterized entanglement \textit{after} classical data loading acts as an unstructured phase-scrambler. Whereas the Untrained Adjoint Pre-Embedding model maintains well-defined geometric coordinates (routinely outperforming Classical DTW before any training occurs), the Untrained Local Observable model collapses to mathematical randomness. This structural degradation is immediately evident on datasets like \textit{RacketSports} ($C=1, 6$) and \textit{Apnea} ($C=2$), where untrained accuracy immediately plummets to the random Zero-R guessing floors (24.59\% and 50.00\%, respectively).

Second, the $\langle Z_0 \rangle$ measurement creates a severe structural bottleneck (quantum pooling). Compressing the geometric overlap of a high-dimensional multivariate state into the scalar expectation value of a single qubit forces the architecture to discard vital Hilbert space information. While the optimizer occasionally finds brittle, highly over-fitted local minima (e.g., 100.00\% on \textit{BasicMotions} $C=6$), the architecture fails catastrophically on generalized real-world sequence alignment. On the \textit{Apnea} dataset ($C=2$), the trained Local Observable model peaks at a mere 65.00\%, vastly trailing both the Adjoint model (90.62\%) and Classical DTW (89.38\%).

The magnitude of this bottleneck failure is fully exposed by the $C=8$ \textit{FingerMovements} dataset. Attempting to pool 8 channels of complex classical variance into the measurement of Qubit 0 completely paralyzes the Triplet Loss optimization. Rather than separating the classes, training the Local Observable model actively degrades its performance, dropping from an untrained 58.33\% to 54.76\%, more than 10\% lower than the Adjoint model (65.48\%). This empirically validates that multivariate time-series alignment strictly requires the global spatial mapping provided by the Adjoint fidelity metric, as localized pooling cannot support high-dimensional data correlation

\section{Discussion}
\label{sec:discussion}

In this work, we introduced a Unified Pre-Embedding Ansatz to establish a robust and scalable quantum distance metric for Dynamic Time Warping (qDTW). Through rigorous empirical ablation across varying dimensionalities ($C=1$ to $C=8$) and distinct data morphologies (kinematic versus physiological), our results demonstrate that quantum sequence alignment can routinely match or exceed classical dynamic programming baselines. However, realizing this state-of-the-art performance required systematically identifying and overcoming several fundamental topological hazards inherent to parameterized quantum circuits. In the following sections, we interpret the physical mechanisms governing these empirical outcomes, synthesize the core structural rules for quantum temporal alignment, and outline the hardware considerations for future architectures.

\subsection{The Spatial-Temporal Tradeoff in Quantum Sequence Alignment}

The primary theoretical contribution of this work lies in identifying the fundamental expressivity tradeoff between spatial width and temporal depth in quantum sequence alignment. Our ablation studies reveal that the optimal quantum architecture is strictly dictated by the classical dimensionality of the input data.

In dimensionally restricted regimes (univariate time series, $C=1$), a single-qubit register fundamentally lacks the spatial degrees of freedom required to map complex decision boundaries. In this scenario, temporal depth injected via classical data re-uploading is strictly necessary. By allowing the single qubit to iteratively twist its basis state around multiplicatively scaled data, the architecture artificially inflates its spatial volume, successfully surpassing classical DTW baselines. 

However, translating this temporal paradigm to multivariate sequences ($C>1$) triggers a catastrophic optimization failure. A multi-qubit register natively possesses an exponentially large spatial volume ($2^C$ dimensions). When this massive spatial capacity is combined with deep temporal re-uploading, the resulting Fourier frequency-spectrum explosion \cite{schuld2021effect} over-parameterizes the model. The quantum phase wraps chaotically, paralyzing the optimizer and completely stalling the Triplet Margin Loss.

This establishes a critical architectural rule for quantum metric learning: spatial width (entanglement) and temporal depth (re-uploading) function as mutually exclusive expressivity mechanisms. Because the classical dynamic programming matrix natively handles temporal sequence processing, the underlying multivariate quantum circuit must act strictly as a stable spatial coordinate system. By physically isolating the classical data from the trainable entanglement, our single-load Pre-Embedding Ansatz perfectly satisfies this requirement, preventing optimization paralysis while dominating across high-dimensional feature spaces.

\subsection{The Supremacy of Adjoint Pre-Embedding and Global Fidelity}
\label{subsec:discussion_adjoint}

Beyond the spatial-temporal tradeoff, our results definitively prove that multivariate sequence alignment strictly requires a global spatial mapping, invalidating the use of traditional local observables ($\langle Z_0 \rangle$). 

The severe degradation observed in the Local Observable ablation highlights a fundamental information bottleneck in quantum pooling. When $C$ channels of classical data are embedded into a $C$-qubit register, the parameterized operations distribute the sequence correlations across an exponentially large, $2^C$-dimensional Hilbert space. Reading out this highly entangled, multi-dimensional state through the scalar expectation value of a single designated qubit inevitably discards vital spatial geometry. On the complex $C=8$ \textit{FingerMovements} dataset, this bottleneck completely derailed the optimizer, actively degrading accuracy from 58.33\% to 54.76\% during training. Conversely, the Adjoint fidelity metric ($\langle \psi_A | \psi_B \rangle$) computes a global geometric overlap by projecting the entire multi-qubit register back onto the ground state. This allows the qDTW matrix to utilize the full mathematical bandwidth of the Hilbert space to calculate point-to-point distances, completely bypassing the quantum pooling bottleneck to hit peak performance.

Furthermore, the architectural order of operations dictates the stability of the latent coordinate system. Specifically, the placement of the trainable entanglement relative to the classical data is crucial. When trainable entanglement is applied \textit{after} the data is loaded (post-embedding), the quantum gates act as unstructured phase-scramblers. The initial relative distances between classical sequences are destroyed before they can be measured, causing the untrained model to collapse to random guessing (e.g., 24.59\% on \textit{RacketSports} and 50.00\% on \textit{Apnea}). 

Our Unified Pre-Embedding Ansatz cures this topological hazard by completely decoupling the parameters from the data. By applying the parameterized layers first, the circuit establishes a stable, high-dimensional coordinate system. The classical sequence is then monotonically loaded onto this fixed manifold. This architecture inherently preserves the native topology of the data, generating well-separated latent geometries that routinely outperform classical Euclidean metrics even before a single gradient update is applied. Thus, we conclude that pre-embedding basis preparation combined with global Adjoint measurement is not merely an architectural alternative, but a strict topological requirement for high-dimensional quantum sequence alignment.

\subsection{The Computational Tradeoff: Untrained Kernels vs. Parameterized Optimization}
\label{subsec:computation_tradeoff}

A defining advantage of classical DTW coupled with a $k$-NN classifier is its status as a non-parametric, "lazy learning" algorithm. By avoiding the resource-intensive gradient optimization phases required by deep neural networks (e.g., CNNs, LSTMs, or Transformers), classical DTW provides immediate inference. Introducing Triplet Loss optimization to the quantum metric fundamentally alters this paradigm, imposing a significant computational training overhead. Consequently, a critical evaluation must be made regarding whether the accuracy improvements of the trained qDTW model justify this physical and computational cost.

A comprehensive analysis of the dimensional scaling data (Table \ref{tab:main_results}) reveals a highly dataset-dependent tradeoff. For certain sequence topologies, gradient-based optimization yields negligible or even negative predictive gains. On the \textit{RacketSports} dataset, the Untrained qDTW baseline consistently matches or slightly outperforms the Trained qDTW model across nearly all channel configurations (e.g., 85.25\% vs. 81.97\% at $C=4$). Similarly, on the \textit{Apnea} dataset ($C=2$), the Untrained architecture establishes the absolute global peak of 92.50\%, while training actively degrades performance to 90.62\%. In these regimes, the single-load Pre-Embedding Ansatz deterministically projects the classical data into a highly expressive Hilbert space where the initial random geometry is naturally separable. Here, the Untrained qDTW emerges as the optimal architectural choice, preserving the "lazy learning" computational benefits of classical DTW while simultaneously upgrading the underlying spatial metric.

Conversely, the necessity of the parameterized training phase becomes absolute when navigating highly entangled, multi-channel kinematic data where random projections fail. The \textit{BasicMotions} dataset perfectly illustrates this physical threshold. At lower dimensions ($C \le 2$), the Untrained kernel effortlessly handles the feature space. However, as dimensionality scales ($C \ge 4$), the dense kinematic overlap causes both Classical DTW and the Untrained qDTW kernel to collapse to 75.00\%. In this chaotic topology, gradient-based optimization is strictly required. By dynamically twisting the quantum basis state to actively separate overlapping class manifolds, the Trained qDTW model recovers massive performance margins, dominating with 93.75\% at $C=4$ and 87.50\% at $C=6$. 

The \textit{FingerMovements} dataset corroborates this necessity at the highest dimensional limits. Across the majority of the spatial scaling ($C=3, 4, 6, 8$), parameterized training systematically untangles the cross-channel variance, providing a consistent 3\% to 7\% accuracy boost over the Untrained projection, ultimately securing the peak performance (65.48\%) at full $C=8$ integration.

Ultimately, this empirical data dictates a bifurcated deployment strategy for quantum sequence alignment. The computationally free Untrained Ansatz should serve as the powerful, default universal baseline, as its initial geometric projection is frequently sufficient to rival classical dynamic programming. Expensive parameterized optimization should be treated as a specialized architectural mechanism, reserved strictly for hyper-dimensional, highly overlapping sequence topologies where initial quantum projections cannot natively separate the feature space.

\subsection{Hardware Considerations and Scaling Limits in the NISQ Era}
\label{subsec:hardware_scaling}

It is crucial to contextualize the performance gains observed in this study by distinguishing between the limitations of our classical simulations and the theoretical scaling of the proposed architecture. Because all experiments were executed using exact, noiseless state-vector simulators, the superiority of the qDTW architecture over classical baselines does not presently stem from physical quantum execution speedup. Rather, the advantage is entirely a product of the quantum metric's profound mathematical expressivity. The architecture leverages the native topology of the Hilbert space and the periodic boundaries of rotational gates to impose a highly structured, geometry-preserving inductive bias on the time-series data, separating complex temporal features much more efficiently than standard Euclidean metrics.

However, simulating the $N \times M$ dynamic programming matrices required for qDTW scales exponentially in classical memory $\mathcal{O}(2^C)$. This computational bottleneck explicitly capped our empirical validation at the $C=8$ dimensionality of the \textit{FingerMovements} dataset. On native quantum hardware, this bottleneck vanishes, as the proposed Pre-Embedding Ansatz scales linearly $\mathcal{O}(C)$ in both qubit allocation and spatial gate operations. Thus, the underlying mathematical framework remains fully scalable to hyper-dimensional time series in the fault-tolerant era.

Despite this theoretical scaling advantage, deploying this mathematical expressivity onto physical Noisy Intermediate-Scale Quantum (NISQ) hardware introduces severe practical limitations. First, the Pre-Embedding Adjoint Ansatz relies on an unperturbed global geometric overlap ($\langle \psi_A | \psi_B \rangle$). On physical Quantum Processing Units (QPUs), shot noise, finite sampling, and hardware-specific decoherence will inherently blur this fidelity measurement. Furthermore, while $L=3$ prevents barren plateaus theoretically, compiling deep cascading CNOT rings on limited qubit connectivity topologies (e.g., heavy-hex lattices) could introduce significant error rates, potentially degrading the coordinate geometries observed in simulation.

Second, the transition from classical simulation to physical hardware introduces a massive optimization bottleneck. In our simulations, gradient descent is efficiently accelerated via classical backpropagation. On a physical QPU, backpropagation is prohibited by measurement collapse and the no-cloning theorem. Consequently, hardware gradients must be extracted iteratively using techniques such as the Parameter-Shift Rule \cite{crooks2019gradients, mitarai2018quantum}. When applied to the recursive evaluations of a dynamic programming matrix, the total number of physical circuit executions required for a single training epoch becomes prohibitively expensive.

This physical training bottleneck profoundly elevates the significance of the Untrained qDTW results presented in this study. Because the single-load Pre-Embedding Ansatz natively clusters temporal features via its mathematical inductive bias, frequently achieving state-of-the-art classification accuracies without requiring a single gradient update, it bypasses the hardware optimization bottleneck entirely. This establishes the untrained Unified Ansatz not merely as a theoretical baseline, but as a highly practical, deployment-ready algorithm for NISQ devices where parameter optimization remains physically intractable.

\subsection{Future Work: Towards Quantum Temporal Memory}

While the Pre-Embedding Ansatz presented in this work successfully stabilizes pointwise distance calculations for qDTW, it treats each time step as an independent spatial event. A promising avenue for future research involves adapting this architecture to ingest sliding temporal windows ($x_t, x_{t+1}, \dots, x_{t+k}$) sequentially through the parameterized layers, effectively functioning as a Quantum Temporal Convolutional Network (QTCN). Such an approach could embed localized temporal memory directly into the quantum phase. However, deploying sequential embeddings within an Adjoint Fidelity framework presents significant open challenges; specifically, the design of novel entanglement topologies required to prevent approximate unitary cancellation ($S^\dagger S \approx I \rightarrow V^\dagger V \approx I$) during highly collinear or resting-state sequence segments.

\section{Conclusion}
\label{sec:conclusion}

In this work, we introduced the Unified Pre-Embedding Ansatz, establishing a highly scalable and mathematically robust quantum distance metric for Dynamic Time Warping (qDTW). By physically decoupling the parameterized entanglement from the classical data encoding, our architecture effectively maps complex multivariate time-series features into the exponentially large quantum Hilbert space. Evaluated across diverse kinematic and physiological datasets scaling up to $C=8$ spatial dimensions, the proposed qDTW models successfully mitigated the classical curse of dimensionality. The architecture routinely matched or exceeded the discriminative performance of standard dynamic programming baselines, leveraging both the native geometric expressivity of the untrained quantum hypersphere and the targeted spatial separation power of Triplet Margin Loss optimization.

Beyond empirical performance, our structural ablations identified fundamental topological laws governing quantum sequence alignment. First, we identified a critical expressivity-trainability threshold, demonstrating the compromise between the circuit depth necessary to untangle multivariate data while strictly avoiding the barren plateaus and representation collapse observed in deeper parameterizations. We also established a strict spatial-temporal tradeoff: while temporal depth (data re-uploading) is a necessary compensatory mechanism to inflate the expressivity of dimensionally restricted single-qubit circuits, applying it to wide multi-qubit registers triggers a frequency-spectrum explosion that causes catastrophic optimization paralysis. Finally, we empirically validated that high-dimensional temporal alignment strictly requires a global spatial mapping via the Adjoint fidelity metric, as traditional local observables ($\langle Z_0 \rangle$) force severe information bottlenecks and induce unstructured phase-scrambling.

Ultimately, this study proves that parameterized quantum circuits do not require deep temporal interleaving to excel at sequence classification, provided the underlying classical dynamic programming matrix handles the temporal alignment. Furthermore, our findings dictate a bifurcated computational strategy for quantum deployment: utilizing the computationally free untrained ansatz as a highly capable, "lazy learning" universal baseline, while reserving resource-intensive gradient optimization strictly for deeply overlapping, hyper-dimensional topologies where classical Euclidean metrics fail. By utilizing the native geometric inductive bias of the quantum Hilbert space to construct a stable coordinate system, the Pre-Embedding Ansatz establishes a rigorous architectural standard for multivariate quantum time-series analysis and provides a foundation for future sequential memory models in the quantum machine learning domain.

\section{Acknowledgment}
This study has been supported by project \text{RYC2022-038121-I}, funded by \lword{MCIN/AEI/10.13039/501100011033} and European Social Fund Plus (ESF+), project \text{PID2023-147422OB-I00} funded by \lword{MCIU/AEI/10.13039/501100011033} and the European FEDER program, and by project \text{ED431F 2025/35} funded by Xunta de Galicia.
The Sleep Heart Health Study (SHHS) was supported by National Heart, Lung, and Blood Institute cooperative agreements U01HL53916 (University of California, Davis), U01HL53931 (New York University), U01HL53934 (University of Minnesota), U01HL53937 and U01HL64360 (Johns Hopkins University), U01HL53938 (University of Arizona), U01HL53940 (University of Washington), U01HL53941 (Boston University), and U01HL63463 (Case Western Reserve University). The National Sleep Research Resource was supported by the National Heart, Lung, and Blood Institute (R24 HL114473, 75N92019R002).

\bibliographystyle{ieeetr}
\bibliography{references}  






\newpage

\appendix

\section{Appendix: Ablation Study on Circuit Depth: Balancing Expressivity and Trainability}
\label{app:ablation_study_circuit_depth}

To determine the optimal temporal depth for the quantum feature map, an ablation study was conducted evaluating circuit depths of $L \in \{1, 3, 5\}$. The primary objective was to identify the optimal architectural compromise between spatial expressivity, i.e., the ability of the parameterized quantum circuit to entangle cross-channel features, and empirical trainability, which is fundamentally limited by the topology of the optimization landscape in deep quantum models. 

\begin{figure}[htbp]
    \centering
    \includegraphics[width=0.9\textwidth]{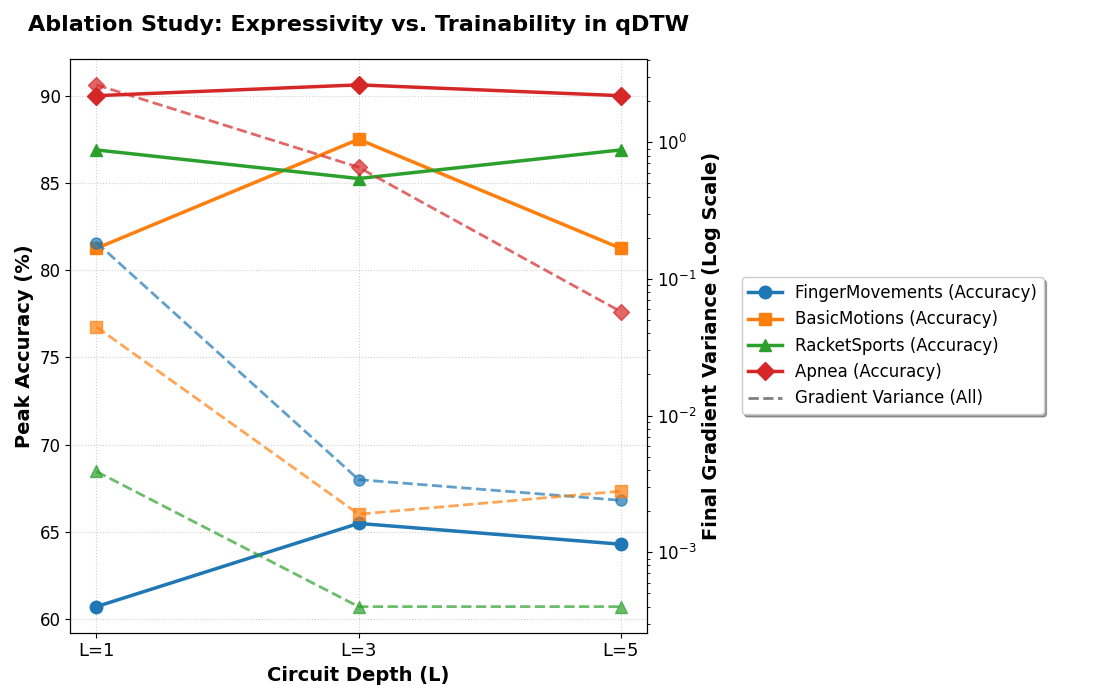}
    \caption{Ablation study evaluating circuit depth ($L \in \{1, 3, 5\}$) across four multivariate time-series datasets. The primary y-axis (solid lines) displays peak classification accuracy, while the secondary y-axis (dashed lines, log scale) indicates the final gradient variance. The data demonstrates that $L=3$ provides the optimal balance between spatial expressivity and optimization stability, whereas $L=5$ triggers a barren plateau and representation collapse in the majority of datasets.}
    \label{fig:ablation_study}
\end{figure}

This evaluation was performed across four distinct multivariate time-series datasets (\textit{BasicMotions}, \textit{RacketSports}, \textit{Apnea}, and \textit{FingerMovements}). Figure~\ref{fig:ablation_study} illustrates the results of this study via a dual-axis plot, contrasting the peak classification accuracy against the final gradient variance observed at the end of the training epochs for each configuration.

The empirical data reveals three distinct phases of model behavior governed by circuit depth:

\subsection{The Expressivity Bottleneck ($L=1$)}
At a minimal depth of $L=1$, the quantum circuit demonstrates highly stable gradients, maintaining relatively high variance (e.g., 0.1848 for \textit{FingerMovements} and 2.6329 for \textit{Apnea}). This stability allows for rapid initial convergence. However, the architecture is fundamentally under-parameterized for complex multivariate tasks. The limited gate count restricts the spatial entanglement necessary to capture multi-channel dependencies. Consequently, the Triplet Loss optimization struggles to carve out sharp decision boundaries, leading to diffuse intra-class clusters and sub-optimal peak accuracies in three of the four datasets.

\subsection{Barren Plateaus and Representation Collapse ($L=5$)}
Conversely, expanding the circuit depth to $L=5$ demonstrates severe diminishing returns, illustrating a classic Quantum Machine Learning trainability failure. As the parameter space expands, the optimization landscape flattens significantly. As shown in Figure~\ref{fig:ablation_study}, the final gradient variance plummets across all datasets (e.g., dropping to 0.0024 in \textit{FingerMovements} and 0.0577 in \textit{Apnea}). Trapped on this barren plateau, the optimizer minimizes the loss function not by maximizing the inter-class margin, but through representation collapse---contracting the entire embedding space toward the origin. This spatial collapse causes class boundaries to overlap, systematically degrading classification accuracy compared to shallower networks.

\subsection{The Architectural Optimum ($L=3$)}
Across the evaluated datasets, $L=3$ emerges as the most robust generalized baseline. Expanding the circuit to this depth provides the precise mathematical flexibility required to overcome the expressivity bottleneck of $L=1$. It introduces sufficient parameterized gates to untangle cross-channel data and map distinct class distributions into separable regions of the Hilbert space. This is evidenced by $L=3$ achieving the absolute highest accuracy for \textit{BasicMotions} (87.50\%), \textit{FingerMovements} (65.48\%), and \textit{Apnea} (90.62\%). 

While gradient variance does decay at $L=3$, it remains just high enough for the optimizer to navigate the loss landscape effectively before the total collapse observed at $L=5$. The sole exception is the \textit{RacketSports} dataset, which sees a marginal accuracy dip at $L=3$ (85.25\%) compared to $L=1$ (86.89\%). This indicates that for datasets with simpler spatial distributions, shallower circuits are adequate, and adding depth triggers premature gradient collapse (variance dropping to 0.0004). 

\subsection{Conclusion}
The results demonstrate that $L=3$ provides the maximum spatial expressivity before the optimization landscape flattens into an untrainable plateau. Therefore, to ensure that the proposed qDTW architecture is evaluated at its peak representational capacity without being bottlenecked by optimization paralysis, $L=3$ is selected as the standard circuit depth for all subsequent primary experiments.

\section{Appendix: Impact of Loss Topology on Quantum Embedding Stability}
\label{app:ablation_loss}

To evaluate the interaction between the optimization landscape and the periodic boundaries of the quantum Hilbert space, an ablation study was conducted comparing Pairwise Contrastive Loss and Triplet Margin Loss across varying margin thresholds ($\beta \in \{0.01, 0.1\}$). As demonstrated in Table \ref{tab:results2}, the empirical results indicate that there is no universally superior loss function; rather, performance is dictated by a fundamental trade-off between absolute separation power and topological stability.

\begin{table}[h]
\centering
\caption{Peak $k$-NN Classification Accuracies Across Datasets with the trained qDTW model under different base margins $\beta$ = \{0.1, 0.01\}, and Triplet (T) or Pairwise (P) loss}
\label{tab:results2}
\begin{tabular}{l c c c c}
\hline
\textbf{Dataset} & \textbf{0.01 T} & \textbf{0.1 T} & \textbf{0.01 P} & \textbf{0.1 P} \\
\hline
BasicMotions & \textbf{87.50\%} & \textbf{87.50\%} & 68.75\% & \textbf{87.50\%} \\
RacketSports & \textbf{85.25\%} & 83.61\% & 75.41\% & 77.05\% \\
FingerMovements & 65.48\% & 60.71\% & \textbf{67.86\%} & \textbf{67.86\%} \\
Apnea & 90.62\% & 91.25\% & 90.62\% & \textbf{91.88\%} \\
\hline
\end{tabular}
\end{table}

Triplet Loss emerges as the highly stable generalist. By enforcing a strictly relative spatial hierarchy, ensuring that an anchor sequence is closer to a positive match than a negative one, it allows the optimizer to separate classes without aggressively pushing quantum states toward the periodic limits of the rotational gates. This topological safety yields dominant performance on \textit{RacketSports} (outperforming Pairwise by nearly 10\% at $\beta=0.01$) and ensures perfect stability across margin settings on \textit{BasicMotions} (maintaining 87.50\% accuracy).

Conversely, Pairwise Loss acts as a high-risk, high-reward specialist. Because it enforces absolute spatial boundaries, it is highly susceptible to phase-wrapping on a closed quantum hypersphere. If an absolute margin pushes the parameterized states too far, it paradoxically decreases the geometric distance between disparate classes, leading to the brittle, catastrophic collapse observed on \textit{BasicMotions} (dropping to 68.75\% at $\beta=0.01$).

However, the analysis of the highly complex 8-channel \textit{FingerMovements} dataset reveals the exact scenario where absolute boundaries are required. On this dataset, Pairwise optimization achieved the peak embedding (67.86\%), whereas Triplet Loss underperformed (65.48\%) and actively degraded when the margin was widened (60.71\%). This indicates that for exceptionally entangled, low-margin feature spaces, the relative clustering pressure of Triplet Loss is insufficient to disentangle the latent space. In such scenarios, the rigid, absolute constraints of Pairwise optimization successfully force the necessary cluster separation. Ultimately, the selection of the loss function in quantum metric learning must be treated as a highly dataset-dependent hyperparameter, balancing the necessity for aggressive separation against the risk of topological collapse.

Despite these dataset-dependent nuances, establishing a standardized optimization baseline is necessary to evaluate the generalized capabilities of the proposed qDTW architecture. In the context of a universal quantum feature map, prioritizing topological safety and consistent convergence supersedes dataset-specific absolute margin tuning. Because Triplet Margin Loss effectively mitigates the risk of catastrophic phase-wrapping and provides the most stable performance across the majority of evaluated sequence topologies, it is selected as the default objective function. Furthermore, the empirical data dictates the selection of a tight margin multiplier ($\beta = 0.01$). While a wider margin ($\beta = 0.1$) yielded marginal gains on simpler topologies like \textit{Apnea}, it actively degraded performance on highly entangled datasets like \textit{FingerMovements} (dropping from 65.48\% to 60.71\%). A tighter relative margin ensures that the optimizer creates distinct intra-class clusters without aggressively stretching the embedding space and risking spatial distortion on the bounded Bloch sphere. Consequently, Triplet Margin Loss with $\beta=0.01$ serves as the foundation for all subsequent primary experiments.

\section{Appendix. Scalable Hybrid Implementation Details}
\label{app:implementation}

The integration of dynamic programming algorithms with parameterized quantum circuits introduces severe computational and memory bottlenecks. A naive Quantum DTW implementation requires evaluating the quantum state for every cell in the $N \times M$ cost matrix, leading to unmanageable sequential execution overhead. To scale the architecture to tens of thousands of time-series evaluations without hardware failure or gradient corruption, our PyTorch training engine utilizes three critical engineering optimizations.

\subsection{Batched Pre-Computation of the Warping Band}
To eliminate sequential quantum simulation overhead, our implementation decouples the point-to-point quantum distance calculations from the recursive DTW pathing algorithm. Prior to matrix construction, the engine pre-calculates all valid $(i, j)$ index pairs permitted by the Sakoe-Chiba band constraint ($W$). The corresponding feature vectors are stacked into batched PyTorch tensors. This allows the state-vector simulator to execute the Parameterized Quantum Circuit (PQC) exactly once per sequence comparison, broadcasting the trainable weights across the batch and evaluating all pointwise quantum distances simultaneously in parallel. Furthermore, due to the PCIe bus transfer overhead associated with small quantum state vectors, we found that forcing execution on native CPU architectures was significantly faster than GPU offloading for this specific PQC topology.

\subsection{Dual-Path Autograd Architecture}
Training a dynamic programming matrix via backpropagation frequently triggers inplace operational errors within standard automatic differentiation engines, as the recursive accumulation matrix is constantly overwritten. To circumvent this without losing the computational graph, we developed a dual-path execution strategy:

\begin{itemize}
    \item \textbf{Training Path (Autograd Active):} During the forward training pass, the dynamic programming matrix is populated using native Python lists of scalar PyTorch tensors rather than a single multi-dimensional tensor. This completely prevents inplace tensor modifications and allows the PyTorch autograd graph to cleanly track the chained $\min()$ operations necessary for backpropagation.
    \item \textbf{Inference Path (Evaluation Mode):} During evaluation and inference, the tracking of gradients is strictly disabled. The batched quantum distances are instantly detached from the PyTorch graph and converted into native NumPy arrays. The Sakoe-Chiba constrained path is then executed using highly optimized standard CPU instructions. By replacing PyTorch stack operations with native Python \texttt{min()} functions on scalar arrays, the recursive evaluation bottleneck is eliminated.
\end{itemize}

\subsection{Optimization Stabilization}
Because quantum parameters $\theta$ map to physical rotations on the Bloch sphere, standard gradient descent can cause violent phase-wrapping if updates are too large. To stabilize the Triplet Loss optimization landscape, we explicitly applied gradient clipping (\texttt{max\_norm = 1.0}) to prevent erratic optimizer jumps, alongside L2 weight decay (\texttt{1e-4}) to act as a regularizing ``spring'' on the parameterized gates.

\section{Appendix: Code and Data Availability}
The complete PyTorch and PennyLane source code utilized to generate the datasets, execute the quantum simulations, and reproduce the ablation studies presented in this manuscript is open-source and publicly available. The repository includes the fully vectorized qDTW implementation and the dynamic programming dual-path autograd engine. The code and instructions for reproducibility can be accessed via GitHub at: \url{https://github.com/diegoalvareze/qDTW}. 

\noindent \textit{Note: The public UEA datasets are automatically fetched via the provided pipeline. The clinical Apnea dataset (derived from SHHS) is subject to strict physiological data privacy agreements (DAUA). Raw polysomnography data cannot be redistributed but can be requested by researchers directly through the National Sleep Research Resource (NSRR) at \url{https://sleepdata.org}.}

\end{document}